

\magnification=1200
\ \bigskip
\centerline{{\bf Random hydrophilic-hydrophobic copolymers}}
\medskip
\centerline{by}
\medskip
\centerline{{\bf T. Garel}}
\centerline{{\sl CEA, Service de Physique Th\'eorique\/}\footnote {$ ^{\dagger}
$}{Laboratoire de la Direction des Sciences de la Mati\`ere du Commissariat
\`a l'Energie Atomique}}
\centerline{{\sl CE-Saclay, 91191 Gif-sur-Yvette Cedex, France\/}}
\medskip
\centerline {{\bf L. Leibler}}
\medskip
\centerline{{\sl Groupe de Physico-Chimie Th\'eorique\/}\footnote {$ ^{\dagger
\dagger}$}
{URA CNRS 1382}}
\centerline{{\sl E.S.P.C.I., 10 rue Vauquelin,}}
\centerline{{\sl 75231 Paris Cedex 05, France }}
\medskip
\centerline{and}
\medskip
\centerline{{\bf H. Orland} \footnote{*}
{Also at Groupe de Physique Statistique, Universit\'e de
Cergy-Pontoise, 95806 Cergy-Pontoise Cedex, France.}}
\medskip
\centerline{{\sl CEA, Service de Physique Th\'eorique\/}$ ^{\dagger}
$}
\centerline{{\sl CE-Saclay, 91191 Gif-sur-Yvette Cedex, France\/}}
\noindent \vglue4truecm \par
\smallskip
\smallskip
\noindent \vglue 4truecm \par
\noindent Submitted for publication to \par
\noindent \lq Journal de Physique \lq
\rq\rq\ \hfill{Saclay, SPhT/94-077}
\vfill\eject

\baselineskip=24pt
\noindent{\bf Abstract} \par
\smallskip
\par
We study a single statistical amphiphilic copolymer chain AB in a selective
solvent
(e.g.water).
Two situations are considered.
In the annealed case, hydrophilic (A) and hydrophobic (B) monomers
are at local chemical
equilibrium and both the fraction of A monomers and their
location along the chain can vary, whereas in
the quenched case (which is relevant to
proteins), the chemical sequence along the chain is fixed by synthesis.
In both cases,
the physical behaviour depends on the
average hydrophobicity of the
polymer chain. For a strongly hydrophobic chain (large fraction of B),
we find an ordinary continuous $\theta$ collapse, with
a large conformational entropy in the collapsed phase.
For a weakly hydrophobic, or a hydrophilic chain, there is an unusual
first-order
collapse transition. In particular, for the case of Gaussian disorder,
this discontinuous transition is driven by
a change of sign of the third virial coefficient.
The entropy of this collapsed phase is strongly reduced with respect
to the $\theta$ collapsed phase.
\vfill\eject
\centerline{{\bf Copolym\`eres al\'eatoires hydrophiles-hydrophobes}}
\medskip
\par
\noindent{\bf R\'esum\'e} \par
\smallskip
\par
Nous \'etudions un copolym\`ere al\'eatoire amphiphile AB
dans un solvant s\'electif (par exemple, de l'eau).
Nous consid\'erons
deux cas. Dans le cas du d\'esordre mobile, les monom\`eres
hydrophiles (A) et hydrophobes (B) sont \`a l'\'equilibre chimique
local, et la fraction de monom\`eres A ainsi que leur position dans
l'espace peuvent varier, alors que dans le cas du d\'esordre gel\'e
(qui est reli\'e au probl\`eme des prot\'eines), la s\'equence
chimique
est fix\'ee par synth\`ese. Dans les deux cas, le comportement de la
cha\^\i ne depend de son hydrophobicit\'e moyenne. Pour une cha\^\i ne
fortement hydrophobe (grande fraction de B), on trouve un point
d'effondrement $\theta$ continu ordinaire, avec une grande entropie
conformationnelle. Pour une cha\^\i ne faiblement hydrophobe ou
hydrophile, on trouve une transition inhabituelle du premier ordre.
En
particuler, dans le cas du d\'esordre gaussien, cette transition
discontinue est pilot\'ee par un changement de signe du troisi\`eme
coefficient du viriel. L'entropie de cette phase collaps\'ee est
fortement r\'eduite par rapport \`a celle d'un point $\theta$
ordinaire.
\vfill\eject
\noindent{\bf I. Introduction} \par
\smallskip
The study of random heteropolymers in solutions has been recently developped,
the
main reason being a plausible
connection of this problem with the protein folding puzzle $^{(1,2)}$.
Various kinds of quenched randomness have been considered. Here, we wish to
emphasize the role of the solvent (water) in the folding process.
Indeed, it is widely believed that the hydrophobic effect $^{(3,4)}$ is
the main driving force for the folding transition: in a native protein,
the hydrophobic residues are burried in an inner core, surrounded by
hydrophilic (or less hydrophobic) residues. In this paper,
we
consider a simple model, where the monomers of a single chain are
(quenched) randomly hydrophilic or hydrophobic.
They interact
with the solvent molecules through an effective
short range interaction,
which is attractive or repulsive, depending on their nature.
A related model
has been considered by Obukhov $^{(5)}$,
but he did not study its phase diagram.
We first consider
the corresponding annealed case, which is simpler and gives useful insights
into the nature of the low temperature phase.
It may well be of interest for
hydrophilic polymers in the presence of complexing amphiphilic
molecules.
In section II, we describe the model.
Section III deals with the annealed case, and section IV with the
quenched case,
both for Gaussian and binary distributions of hydrophilicities.
\par
\noindent{\bf II. The model} \par
\smallskip
\par
We consider a chain of $N$ monomers which can be either hydrophilic
(L) or hydrophobic (P). The hydrophilicity degree of monomer $i$ is
measured by $\lambda_i$, and the monomer-solvent interaction is assumed to
be short-ranged ($\delta$-interaction). More precisely,
we consider a
monomer-solvent interaction of the form:
$$
{\cal H}_{ms} = - \sum_{i=1}^{N+1} \sum_{\alpha=1}^{{\cal N}} \lambda_i
\delta ( {\vec
r_i}-{\vec R_{\alpha}} )
\eqno (1)
$$
where the ${\vec r}_i$ are the positions of the $N$ monomers and
the ${\vec R}_{\alpha}$ are the positions of the $\cal N$
solvent molecules \footnote{$^{1}$}
{Note that $-\lambda /kT$ is identical to the standard Flory
parameter $\chi$, often used
in the polymer literature.}.
Here, the $\lambda_i$ are independent annealed or
quenched random variables.
A positive ( resp. negative) $\lambda$ corresponds to a hydrophilic
(resp. hydrophobic) monomer. Using the incompressibility
condition of the chain-solvent system $^{(6)}$, i.e. the
sum of the monomer and solvent concentration is constant:
$\rho_m(\vec r)+\rho_s(\vec r) = \rho_0$,
the solvent degrees of freedom can be eliminated, and
the full Hamiltonian reads :
$$
\eqalign{
\beta {\cal H}(\lambda_i) &= {1 \over 2} \sum_{i \ne j} \left [
v_0 + \beta ( \lambda_i+\lambda_j)\right]
 \ \delta(\vec r_i - \vec r_j) \cr
&+ { 1 \over
6 } \sum_{i \ne j \ne k} w_3
\ \delta(\vec r_i- \vec r_j)\ \delta(\vec r_j - \vec r_k) \cr
&+ { 1 \over 24 } \sum_{i \ne j \ne k \ne l} w_4
\ \delta(\vec r_i- \vec r_j)\ \delta(\vec r_j - \vec r_k)
\ \delta(\vec r_k - \vec r_l)
\cr
}
\eqno (2)
$$
In equation (2), the $v_0$ term accounts for entropic excluded volume
interactions as well as effective AB interactions described by Flory
parameter $\chi_{AB}$
\footnote{$^{2}$}
{In the Flory-Huggins model of polymer solutions, $v_0 = a^3
(1-2\chi_{AB})$, $w_3= a^6$ and $w_4=a^9$, where $a$ is the monomer
size$^{(6)}$.}.
Usually, when considering a polymer collapse transition, it is
sufficient to
include terms up to $w_3$.
Here, we have included a repulsive four-body term $w_4$
which will play an essential role.

In the continuous limit $^{(7)}$, the partition function reads:
$$
Z(\lambda(s)) = \int {\cal D} {\vec r(s)} \exp \left( - {d \over 2 a^2
}\int_0^N
 ds
\left( {d {\vec r(s)} \over ds} \right)^2 - \beta{\cal H}(\lambda(s)) \right)
\eqno (3a)
$$
with
$$
\beta {\cal H}(\lambda(s))
= {1 \over 2}\int_0^N ds \int_0^N ds' \left(v_0 + \beta
(\lambda(s)+\lambda(s'))
\right)
\delta({\vec r(s)}-{\vec
r(s')})  $$
$$
+ {w_3 \over 6} \int_0^N ds \int_0^N ds'\int_0^N ds'' \
\delta ({\vec r(s)}-{\vec
r(s')}) \ \delta ({\vec r(s')}-{\vec r(s'')})$$
$$
+ { w_4 \over 24}\int_0^N ds \int_0^N ds'\int_0^N ds'' \int_0^N ds'''
\ \delta({\vec r(s)}-{\vec r(s')})\ \delta({\vec r(s')}-{\vec
r(s'')})
\ \delta({\vec r(s'')}-{\vec r(s''')})
\eqno(3b)
$$
where we have implicitly assumed that the origin of the chain is
pinned
at $\vec 0$ and its extremity is free. The parameter $a$ denotes
the statistical segment (Kuhn) length , and $d$ denotes the dimension of space.
In the following, we shall study the case where
the hydrophilicities $\lambda(s)$ are random independent
Gaussian variables, with average $\lambda_0$ and variance $\Delta$:
$$
P(\lambda_i)= {1 \over \sqrt{2 \pi \Delta^2}} \exp( - {(\lambda_i
-\lambda_0)^2 \over \Delta^2})
\eqno(4)
$$
We shall also discuss briefly the technically
more complicated case where the $\lambda_i$ are
random independent binary
variables, taking the value $\lambda_A$ with probability $p$ for
a hydrophilic monomer and $\lambda_B$ with probability $q=1-p$\ for a
hydrophobic monomer. This case is more relevant to synthetic random copolymers.
\par
The annealed case amounts to take the average of the partition
function $Z$ on the randomness, whereas the quenched case corresponds
to taking the average of $\log Z$. Since the quenched case is achieved
by introducing replicas $^{(8)}$ and using the identity $\log Z = \lim_{n \to
0}
{Z^n -1 \over n}$, we compute directly the average $\overline{Z^n}$
for integer $n$, and then treat
separately the annealed ($n=1$) and quenched ($n=0$) cases.
For Gaussian disorder and integer $n$, the average yields:
$$
\overline{ Z_G^n} =
\int \prod_{a=1}^n {\cal D} {\vec r_a(s)} \exp \left( - {d \over 2 a^2
}
\int_0^N ds \sum_{a=1}^n
\left( {d {\vec r_a(s)} \over ds} \right)^2 - A_G \right)
$$
$$
\exp \left( {\beta^2 \lambda^2 \over 2}
 \int_0^N ds \int_0^N ds'\int_0^N ds'' \
\sum_{a,b=1}^n \delta ({\vec r_a(s)}-{\vec
r_a(s')}) \ \delta ({\vec r_b(s)}-{\vec r_b(s'')}) \right)
\eqno(5a)
$$
with
$$
A_G ={1 \over 2}(v_0 +2 \beta
\lambda_0 ) \int_0^N ds \int_0^N ds' \
\sum_{a=1}^n \delta({\vec r_a(s)}-{\vec r_a(s')})
$$
$$
+ {w_3 \over 6} \int_0^N ds \int_0^N ds'\int_0^N ds'' \
\sum_{a=1}^n \delta ({\vec r_a(s)}-{\vec
r_a(s')}) \ \delta ({\vec r_a(s')}-{\vec r_a(s'')})$$
$$
+ {w_4 \over 24}
\int_0^N ds \int_0^N ds'\int_0^N ds'' \int_0^N ds'''
$$
$$
\sum_{a=1}^n\ \delta({\vec r_a(s)}-{\vec r_a(s')})\ \delta({\vec r_a(s')}-{\vec
r_a(s'')})
\ \delta({\vec r_a(s'')}-{\vec r_a(s''')})
\eqno(5b)
$$
whereas for binary disorder:
$$
\overline{ Z_B^n} =
\int \prod_{a=1}^n {\cal D} {\vec r_a(s)} \exp \left( - {d \over 2 a^2
}
\int_0^N ds \sum_{a=1}^n
\left( {d {\vec r_a(s)} \over ds} \right)^2 - A_B \right)
\eqno(6a)
$$
with
$$
A_B ={v_0 \over 2} \int_0^N ds \int_0^N ds' \
\sum_{a=1}^n \delta({\vec r_a(s)}-{\vec r_a(s')})
$$
$$
+ {w_3 \over 6} \int_0^N ds \int_0^N ds'\int_0^N ds'' \
\sum_{a=1}^n \delta ({\vec r_a(s)}-{\vec
r_a(s')}) \ \delta ({\vec r_a(s')}-{\vec r_a(s'')})
$$
$$
+ {w_4 \over 24}
\int_0^N ds \int_0^N ds'\int_0^N ds'' \int_0^N ds'''
$$
$$
\sum_{a=1}^n\ \delta({\vec r_a(s)}-{\vec r_a(s')})\ \delta({\vec r_a(s')}-{\vec
r_a(s'')})
\ \delta({\vec r_a(s'')}-{\vec r_a(s''')})
$$
$$
- \int_0^N ds \ \log \left( p\ e^{- \beta \lambda_A \sum_{a=1}^n
\int_0^N ds' \delta ({\vec r_a(s)}-{\vec r_a(s')})}
+ q\ e^{ \beta \lambda_B \sum_{a=1}^n
\int_0^N ds' \delta ({\vec r_a(s)}-{\vec r_a(s')})}
\right)
\eqno(6b)
$$
We now study separately the annealed and quenched cases.
\par
\noindent{\bf III. The annealed case} \par
\par
We set $n=1$ in equations (5) and (6).
Denoting the monomer concentration by
$\rho(\vec r)$,
we assume the validity of
ground state dominance $^{(6,9)}$ . The ground state wave-function
$\varphi(\vec r)$ is normalized, and we perform a saddle-point
approximation of the partition function.
We quote here only the results,
leaving the details of the calculations
for the quenched case (section IV).
These approximations yield the following
annealed free energy per monomer:
\itemitem{a)} {\sl Gaussian disorder}
\par
$$
\beta f_G = {a^2\over2d} \int d^d r \left( \nabla \varphi (\vec
r)\right)^2 + {(v_0+2\ \beta\ \lambda_0)\over 2}\ N \ \int d^d r\
\varphi^4(\vec
r)
$$
$$ +N^2 {w'_3 \over 6} \int d^dr\ \varphi^6 (\vec r)
+ {w_4 \over 24} N^3 \int d^dr\ \varphi^8(\vec r)
- E_0 \int d^d r\ \varphi(\vec r)^2
\eqno(7a)
$$
where
$$
w'_3 = w_3- 3 \beta^2 \lambda^2
\eqno(7b)
$$
The energy $E_0$ appearing in (7a) is the Lagrange multiplier enforcing
the normalisation of the wave-function $\varphi$.
The equation of state is obtained by minimizing the free
energy (7a) with respect to the wave-function $\varphi(\vec r)$.
The monomer concentration is given by:
$$
\rho(\vec r) = N \varphi^2 (\vec r)
\eqno(8)
$$
\par
The phase diagram of the system can be easily discussed: for any
non-zero concentration of hydrophobic monomers, there is a
transition between a swollen phase ($\rho = 0$) at high temperature,
and a collapsed phase (finite $\rho$) at low temperature. The nature
of the transition depends on the average hydrophobicity of the chain
$\lambda_0$ and variance $\lambda$.
\par
For strong average hydrophobicity ($\lambda_0 <0$), there is a second
order collapse transition, similar to an ordinary $\theta$ point
for polymers in a bad solvent $^{(10)}$ . This occurs when the coefficient
of the $\varphi^4$ term in (7a) becomes negative while the
coefficient of the $\varphi^6$ term is still positive. This occurs if
the average hydrophobicity $\lambda_0$ satisfies the condition:
$$
- \lambda_0 \ge {v_0 \lambda \over 2}\ \sqrt{{3 \over w_3}}
\eqno(9a)
$$
and the transition temperature is given by:
$$
T_0 = -{2 \lambda_0 \over v_0}
\eqno(9b)
$$
When condition (9a) is not satisfied, the collapse transition becomes
first order, and is driven by the change of sign of $w'_3$, i.e. by a
change of sign of the effective third virial coefficient. This is
quite an unusual mechanism.
\par
Such is the case if the system is weakly hydrophobic, or hydrophilic.
The minimization of the free energy (7a) leads to a non-linear
partial differential equation for $\varphi$ which cannot be solved
analytically. It is possible to restrict the minimization to a
subset of Gaussian wave-functions, and this will be discussed in the
section on quenched systems (see eq.(20) below).
Within this Gaussian approximation,
the first order transition temperature $T^{\star}$ is given by the
equation:
$$
3^{d+1} \ w_4\ (v_0 +{2 \lambda_0 \over T^{\star}})\ =\ 2^{3d/2}\ (w_3 - {3
\lambda^2 \over
{T^{\star}}^ 2} )^2
\eqno(9c)
$$
\par
\itemitem{b)} {\sl Binary disorder}
\par
In this case, the free energy per monomer reads:
$$
\beta f_B =
\int d{\vec r} \left( {a^2\over 2 d} (\nabla \varphi)^2
+N{v_0 \over 2} \varphi^4 + N^2 {w_3 \over 6} \varphi^6 + N^3
{w_4 \over 24} \varphi^8 - G \right)
\eqno(10a)
$$
\par
\noindent
where
$$
G=
\varphi^2 \log\left( p\ e^{- N \beta \lambda_A \varphi^2}
+ q\ e^{ N \beta \lambda_B \varphi^2} \right) + E_0 \varphi^2
\eqno(10b)
$$
By taking derivatives with respect to the chemical potentials $\lambda_{\pm}$,
it is easily seen that the concentrations of hydrophilic and
hydrophobic monomers are given by:
$$
\eqalign{
\rho_+(\vec r) &= N \varphi^2 (\vec r) { p\ e^{- N \beta \lambda_A \varphi^2}
\over  p\ e^{- N \beta \lambda_A \varphi^2}
+ q\ e^{ N \beta \lambda_B \varphi^2} } \cr
\rho_-(\vec r) &= N \varphi^2 (\vec r) { q\ e^{ N \beta \lambda_B \varphi^2}
\over  p\ e^{- N \beta \lambda_A \varphi^2}
+ q\ e^{ N \beta \lambda_B \varphi^2} }
}
\eqno(11)
$$
Note the Boltzman weights in (11), which govern the chemical
composition of the polymer. It clearly appears that the regions of
high monomer concentration are hydrophobic, whereas those of low
concentration are hydrophilic. The main conclusions of the Gaussian
case above remain valid, except that the transition temperatures
cannot be calculated explicitly. We find again an ordinary
second order $\theta$
point for strongly hydrophobic chains, and a first-order collapse for
weakly hydrophobic or hydrophilic chains.
\par
As can be seen by a numerical study of this case, the transition
is strongly first-order, and
this implies the existence of a large
latent heat at the transition,
and thus a sharp drop of the entropy. This drop
in entropy can be interpreted as the formation of a hydrophobic core,
surrounded by a hydrophilic coating, which is energetically
favourable, but entropically unfavourable. This is confirmed by
equations (11), which show that high concentration regions are
hydrophobic, whereas low concentration regions are hydrophilic.
Finally, we see from (11) that at low temperature, the total number
of hydrophilic monomers goes to zero, whereas the number of
hydrophobic monomers goes to $N$.
\par
\noindent {\bf IV. The quenched case}
\par
We have to take the $n=0$ limit in (5) and (6).
Introducing the parameters $q_{ab}(\vec r, \vec r')$ , with $a < b$, and
$\rho_a (\vec r)$ by:
$$
\eqalign{
q_{ab}(\vec r, \vec r') &= \int_0^N ds \ \delta ({\vec r_a(s)}- \vec r)
\ \delta ({\vec r_b(s)}-{\vec r'}) \cr
\rho_a(\vec r)&= \int_0^N ds \ \delta ({\vec r_a(s)}- \vec r) \cr
}
\eqno(12)
$$
we may write equations (5) as:
\itemitem{a)} {\sl Gaussian disorder}
\par
$$
\overline{Z_G^n} = \int {\cal D} q_{ab}(\vec r, \vec r')
{\cal D} {\hat q}_{ab}(\vec r, \vec r') {\cal D} \rho_a(\vec r)
{\cal D} \phi_a(\vec r)
\ \exp \left( G(q_{ab}, {\hat q}_{ab}, \rho_a, \phi_a) + \log
\zeta({\hat q}_{ab}, \phi_a) \right)
\eqno(13a)
$$
where ${\hat q}_{ab}(\vec r, \vec r')$ and $\phi_a (\vec r)$ are the Lagrange
multipliers associated with (12), and:
$$
\eqalign{
G(q_{ab}, {\hat q}_{ab}, \rho_a, \phi_a) &= \int dr \sum_a
\left( i \rho_a(\vec r)
\phi_a(\vec r) - (v_0 + 2 \beta \lambda_0)\ { \rho_a ^2 (\vec r) \over 2}
- {w'_3 \over 6} \rho_a^3 (\vec r) - {w_4 \over 24}
\rho_a^4 (\vec r) \right) \cr
& + \int dr \int dr' \sum_{a < b} \left( i q_{ab}(\vec r, \vec r')
{\hat q}_{ab}(\vec r, \vec r') + \beta^2 \lambda^2 \ q_{ab}(\vec r, \vec r')
\rho_a(\vec r) \rho_b(\vec r') \right) \cr
}
\eqno(13b)
$$
and
$$
\zeta ({\hat q}_{ab}, \phi_a) =
\int \prod_a {\cal D}\vec  r_a(s)$$
$$
\times \exp \left(-{d \over
2 a^2} \int^ N_0  ds\ \sum_a \left({d \vec r_a \over ds}\right)^2 -
i \int_0^N ds\ \sum_a \phi_a ( \vec r_a (s))
-i \int_0^N ds \sum_{a<b}\ {\hat q}_{ab}(\vec r_a(s), \vec r_b (s))
\right)
\eqno(13c)
$$
where $w'_3$ is defined in equation (7b).

Using the identity of equation (13c) with a Feynman integral $^{(9,11)}$
, we can
write:
$$
\zeta ({\hat q}_{ab}, \phi_a) = \int \prod_a d^d r_a
<\vec r_1 \ldots \vec r_n | e^{-N H_n({\hat q}_{ab}, \phi_a)}
 | \vec 0 \ldots \vec 0>
\eqno(14a)
$$
where $H_n$ is a ``quantum-like'' $n \to 0$ Hamiltonian, given by:
$$
H_n= -{ a^2 \over 2 d } \sum_a \nabla_a^2 \ + \ \sum_a i \phi_a(\vec
r_a) + \sum_{a < b} i{\hat q}_{ab}(\vec r_a, \vec r_b)
\eqno(14b)
$$
Anticipating some kind of (hydrophobically-driven) collapse, we assume
that we can use ground-state dominance to evaluate (14), and write,
omitting some non-extensive prefactors:
$$
\eqalign{
\zeta ({\hat q}_{ab}, \phi_a) &\simeq \ e^{-N E_0({\hat q}_{ab},
\phi_a)} \cr
&=\exp\left(-N\min_{\{\Psi(\vec r)\}}\left\{ <\Psi|H_n|\Psi> -
E_0 \left(<\Psi|\Psi>-1 \right)\right\} \right)
}
\eqno(15)
$$
where $E_0$ is the ground state energy of $H_n$.
At this point, the problem is still untractable, and we make the
extra approximation of saddle-point method (SPM). The extremization
with respect to $q_{ab}$ reads:
$$
i{\hat q}_{ab}(\vec r,\vec r') = - \beta^2 \ \lambda^2 \rho_a (\vec
r)\ \rho_b (\vec r')
\eqno(16)
$$
This equation shows that replica symmetry is not broken, since $\hat
q_{ab}$ is a product of two single-replica quantities $\rho_a$.
To get more analytic information, we use the
Rayleigh-Ritz
variational principle to evaluate $E_0$. Due to
the absence of replica symmetry breaking (RSB), we further restrict
the variational wave-function space to Hartree-like
replica-symmetric wave-functions, and write:
$$
\Psi(\vec r_1,\ldots,\vec r_n) = \prod_{a=1}^n \varphi(\vec r_a)
\eqno(17)
$$
Because of replica symmetry, we can omit replica indices and
take easily the $n \to 0$ limit.
The variational free energy now reads:
$$
-\beta \overline{F_G}(q, {\hat q}, \rho, \phi, \varphi) = \int dr
\left( i \rho(\vec r)
\phi(\vec r) - (v_0 + 2 \beta \lambda_0)\ { \rho^2 (\vec r) \over 2}
- {w'_3 \over 6} \rho^3 (\vec r) - {w_4 \over 24}
\rho^4 (\vec r) \right) $$
$$
 -{1\over 2} \int dr \int dr' \left( i q(\vec r, \vec r')
{\hat q}(\vec r, \vec r') + \beta^2 \lambda^2 \ q(\vec r, \vec r')
\rho(\vec r) \rho(\vec r') \right) $$
$$
- N \left\{\int d^d r
\ \varphi (\vec r) \left(-{a^2 \over 2 d } \nabla^2 +  i \ \phi(\vec
r) \ \right) \varphi(\vec r)
 - {i \over 2} \int d^dr \int d^dr'
\hat q(\vec r, \vec r') \varphi^2 (\vec r) \varphi^2 (\vec r') \right\}
$$
$$
+N E_0 \left( \int d^dr \varphi^2(\vec r) -1 \right)
\eqno(18)
$$
The SPM equations read:
$$
\eqalign{
\rho(\vec r) &= N \varphi^2(\vec r) \cr
q(\vec r,\vec r') &= N  \varphi^2(\vec r)  \varphi^2(\vec r') \cr
i\phi(\vec r) &= (v_0 + 2\beta \lambda_0) \rho(\vec r) + {w'_3 \over
2} \rho^2(\vec r) + { w_4 \over 6} \rho^3 ( \vec r) + \beta^2
\lambda^2 \int d^dr' q(\vec r,\vec r') \rho(\vec r') \cr
i \hat q (\vec r,\vec r') &= - \beta^2 \lambda^2 \rho(\vec r) \rho(\vec
r') \cr
}
\eqno(19)
$$
We still have to minimize with respect to the normalized wave-function
$\varphi(\vec r)$. This leads to a very complicated non-linear
Schr\"odinger equation, and we shall restrict ourselves to a
one-parameter family of Gaussian wave-functions of the form:
$$
\varphi(\vec r) =  \left( {1\over 2 \pi R^2} \right)^{d/4}
\exp(- { {\vec r}^2 \over 4 R^2})
\eqno(20)
$$
where $R$ is the only variational parameter.
\par
Using equations (18) and (19),
the variational free energy per monomer ($ \overline{f_G} =
\overline{F_G}/N$) reads:
$$
\beta \overline{f_G} = {a^2\over2d} \int d^d r \left( \nabla \varphi (\vec
r)\right)^2 + {(v_0+2\ \beta\ \lambda_0)\over 2}\ N \
\int d^d r \varphi^4(\vec r)
$$
$$ +N^2 \left \{ {w'_3 \over 6} \int d^dr \varphi^6 (\vec r) +
{\beta^2 \lambda^2 \over 2} \left(\int d^d r \varphi^4 (\vec r) \right)^2
\right \} + {w_4 \over 24} N^3 \int d^dr \varphi^8(\vec r)
$$
$$
-E_0 \int d^dr \varphi^2(\vec r)
\eqno(21)
$$
Note that the only difference between this free energy and the
annealed one of eq.(7a), is the presence, in the quenched case, of an
additional disorder induced term, which we have grouped together
with the $\varphi^6$ term for reasons that will become clear below.
Using the Gaussian wave function (20), and the value of $w'_3$
given in (7b), the free energy reads:
$$
\beta \overline{f_G} = {a^2\over8R^2}+ {1\over (2 \sqrt{\pi})^d}{(v_0+2\
\beta\ \lambda_0)\over 2}{N\over R^d}$$
$$
+ \left( {1\over (2 \pi \sqrt{3})^d}
{w_3\over 6}- {1\over (2 \pi)^d}{\beta^2 \lambda^2 \over
2}(3^{-d/2}-2^{-d}) \right) \  ({N\over R^d})^2 + \left({1\over (32 \pi^3)}
\right)^{d/2}
\  {w_4 \over 24}\  \left({N\over R^d}\right)^3
\eqno(22)
$$
At low temperatures, one has to study  the sign of the third
term of eq(22) . This
sign indeed becomes negative at low temperature, due to disorder
fluctuations, yielding the following results:
\item{i)}{$\lambda_0 > 0$ : the hydrophilic case.}
\par
In our approximation, we find that there is a first order transition
towards a collapsed phase ($R \sim N^{1/d}$) induced by a negative
three-body term (and stabilized by a positive four-body term). This
transition is neither an ordinary $\theta$ point (since the two-body
term is positive), nor a freezing point (the replica-symmetry is not
broken). Note that since our approach is variational, the true free
energy of the system is lower than the variational one, and
we thus expect the real transition to occur at an even higher
temperature. Since the transition is first order, we expect
metastability and retardation
effects to be important near the transition; we also expect that (due
to the latent heat), there will be a reduction of entropy in the low
temperature phase, compared to an ordinary second order $\theta$ point.
\item{ii)}{$\lambda_0 < 0$ : the hydrophobic case.}
\par
In this case, defining the two temperatures:
$$
\eqalign{
T_0&= 2 |\lambda_0| /v_0 \cr
T_1&= \lambda {\sqrt 3 \over \sqrt{w_3(1 - 3^{d/2}/2^d)}} \cr
}
\eqno(23)
$$
leads to two possible scenarii, in a very similar way as in the
annealed case:
\itemitem {a)} Weakly hydrophobic case : $T_1 > T_0$
\par
The collapse transition is again driven
by the disorder fluctuations of the three-body interactions. The
resulting first-order transition is very similar to the hydrophilic
case i).
\itemitem {b)} Strongly hydrophobic case : $T_0 > T_1$
\par
The collapse transition is now driven by
the strong two-body $\lambda_0$ term. The resulting phase transition
is very similar to an ordinary $\theta$ point, and is therefore
second-order.
\par
The phase diagram of the quenched case seems quite similar to the
annealed case, but, as we shall see on the case of binary disorder,
the physics is quite different. Note also that the transition
temperature in the quenched case is lower than in the corresponding
annealed case, as can be checked from (7a) and (21). This may be
viewed as an effect of geometrical frustration induced by the chain
constraint.
\par
\itemitem{b)} {\sl Binary disorder}
\par
Using the same approximations as in the previous case, namely ground
state dominance, no replica symmetry breaking and Hartree variational
wave-function, the free energy per monomer reads:
$$
\beta \overline{f_B} = {a^2\over2d} \int d^d r \left( \nabla \varphi (\vec
r)\right)^2 + {v_0\over 2}\ N \
\int d^d r \varphi^4(\vec r)
$$
$$ +N^2 {w_3 \over 6} \int d^dr \varphi^6 (\vec r) +
{w_4 \over 24} N^3 \int d^dr \varphi^8(\vec r) -E_0 \int d^dr
\varphi^2(\vec r)
$$
$$
+N \beta \lambda_A\int d^dr \varphi^4(\vec r)
- \sum_{l=1}^{\infty} {(-1)^{l+1} \over l} ({q \over p})^l
\ \log \left( \int d^dr \varphi^2(\vec r) e^{N \beta
\lambda l \ \varphi^2(\vec r)} \right)
\eqno(24)
$$
where $\lambda = \lambda_A + \lambda_B$.
\par
In this case, we see that even the Gaussian variational form is not
tractable, but a numerical study shows that the system has exactly
the same phase diagram as for Gaussian disorder. However, the
interest  of this model is that it allows to calculate the
concentration of hydrophilic and hydrophobic monomers. We obtain:
$$
\eqalign{
\rho_+ (\vec r) &= N \varphi^2(\vec r) \left[ 1 - \sum_{l=1}^\infty
(-1)^{l+1}
({q \over p})^l  {e^{\beta \lambda l N \varphi^2} \over \int d^d r
\varphi^2 e^{\beta \lambda l N \varphi^2}} \right] \cr
\rho_- (\vec r) &= N \varphi^2(\vec r) \sum_{l=1}^\infty
(-1)^{l+1}
 ({q \over p})^l  {e^{\beta \lambda l N \varphi^2} \over \int d^d r
\varphi^2 e^{\beta \lambda l N \varphi^2}}  \cr
}
\eqno(25)
$$
It is easily seen on (25) that the two concentrations satisfy the sum rules:
$$
\eqalign{
\int d^d r \rho_+ (\vec r) &= N p \cr
\int d^d r \rho_- (\vec r) &= N q \cr
}
\eqno(26)
$$
and thus (as expected for a quenched chain), the overall chemical
composition  of the chain remains unchanged.
\par
\noindent{\bf V. Conclusion} \par
We have studied a simple model of a randomly hydrophilic-hydrophobic
chain in water. The randomness may be annealed or quenched.
The low temperature
phase is always collapsed; depending
on the degree of hydrophobicity of the chain, this collapsed phase is
reached either through an usual second order $\theta$ point,
or through a first order transition. In all cases (see eq.(11) and (25)),
the interior of the compact globule is mainly hydrophobic, whereas its
exterior is hydrophilic (as expected). In the
case of a first order transition, the existence of a latent heat
implies a decrease of the entropy of the collapsed chain, due to the
burial of the hydrophobic monomers in the core. Studies of similar
models for semi-dilute solutions and melts in random or block
copolymers are currently under way.

\vfill\eject
\bigskip
\centerline{{\bf REFERENCES}}
\medskip
\noindent (1)\/ Karplus M. and Shakhnovich E.I. in
``Protein Folding'' , Creighton T.E. (editor), W.H. Freeman, New
York (1992). \par
\noindent (2)\/ Garel T., Orland H. and Thirumalai D. in
``New Developments in Theoretical Studies of Proteins'', Elber R.
(editor), World Scientific, Singapour (1994). \par
\noindent (3) \/ Tanford C., ``the Hydrophobic effect'', Wiley,
New-York (1980). \par
\noindent (4) \/ Dill K.A. 1990,
{\sl Biochemistry\/} {\bf 29}, 7133. \par
\noindent (5) \/ Obukhov S.P.1986,  {\sl
J.Phys.A\/} {\bf 19}, 3655. \par
\noindent (6) \/de Gennes P.G. 1979, \lq\lq
Scaling concepts in polymer physics\rq\rq , Cornell University Press, Ithaca .
\par
\noindent (7)\/ des Cloizeaux J. and Jannink G. 1987, \lq\lq
Les Polym\`eres en
Solution\rq\rq\ Eds. de Physique, Les Ulis.\par
\noindent (8) \/M\'ezard M., Parisi G. and Virasoro M.A. 1987,
\lq\lq Spin glass theory
and beyond\rq\rq , World Scientific, Singapore.
\par
\noindent (9) \/Wiegel F.W. 1986, \lq\lq
Introduction to path integral methods in physics and polymer
science\rq\rq , World scientific, Singapore.\par
\noindent (10)\/ Flory P.J. 1971, \lq\lq
Principles of polymer chemistry\rq\rq , Cornell University Press, Ithaca.
\par
\noindent (11)\/ Edwards S.F. 1965,  {\sl
Proc.Phys.Soc.London\/} {\bf 85}, 613.\par

\end